\documentclass[a4paper,twoside]{article}

\usepackage{epsfig}
\usepackage{subcaption}
\usepackage{calc}
\usepackage{amssymb}
\usepackage{amstext}
\usepackage{amsmath}
\usepackage{amsthm}
\usepackage{multicol}
\usepackage{pslatex}
\usepackage{apalike}
\usepackage{algorithm2e}
\usepackage[bottom]{footmisc}
\usepackage{SCITEPRESS}     

\usepackage{comment}
\usepackage{amsmath}
\usepackage{tikz}      
\usepackage{quantikz}  
\usepackage{booktabs}  
\usepackage{tabularx}  
\usepackage{enumitem}  
\usepackage{orcidlink}

\begin{document}

\title{Topology-Guided Quantum GANs for Constrained Graph Generation}

\author{\authorname{Tobias Rohe\sup{1}\orcidlink{0009-0003-3283-0586}, Markus Baumann\sup{1}\orcidlink{0009-0007-3575-1006}, Michael Poppel\sup{1}\orcidlink{0009-0005-1141-0974}, Gerhard Stenzel\sup{1}\orcidlink{0009-0009-0280-4911}, Maximilian Zorn\sup{1}\orcidlink{0009-0006-2750-7495}, and Claudia Linnhoff-Popien\sup{1}\orcidlink{0000-0001-6284-9286}}
\affiliation{\sup{1}Mobile and Distributed Systems Group, LMU Munich, Germany}
\email{tobias.rohe@ifi.lmu.de}
}


\abstract{Quantum computing (QC) promises theoretical advantages, benefiting computational problems that would not be efficiently classically simulatable. However, much of this theoretical speedup depends on the quantum circuit design solving the problem. We argue that QC literature has yet to explore more domain specific ansatz-topologies, instead of relying on generic, one-size-fits-all architectures. In this work, we show that incorporating task-specific inductive biases -- specifically geometric priors -- into quantum circuit design can enhance the performance of hybrid Quantum Generative Adversarial Networks (QuGANs) on the task of generating geometrically constrained \(K_4\) graphs. We evaluate a portfolio of entanglement topologies and loss-function designs to assess their impact on both statistical fidelity and compliance with geometric constraints, including the Triangle and Ptolemaic inequalities. Our results show that aligning circuit topology with the underlying problem structure yields substantial benefits: the Triangle-topology QuGAN achieves the highest geometric validity among quantum models and matches the performance of classical Generative Adversarial Networks (GAN). Additionally, we showcase how specific architectural choices, such as entangling gate types, variance regularization and output-scaling govern the trade-off between geometric consistency and distributional accuracy, thus emphasizing the value of structured, task-aware quantum ansatz-topologies.}

\onecolumn \maketitle \normalsize \setcounter{footnote}{0} \vfill

\section{INTRODUCTION}
\label{sec:introduction}

Quantum computing (QC) offers a fundamentally different computational paradigm compared to classical systems, exploiting phenomena such as superposition, interference, and entanglement. These unique quantum resources enable the exploration of high-dimensional Hilbert spaces and the modelling of complex, non-classical correlation patterns. Simulating universal QC on classical hardware is known to be computationally intractable, leading to the widespread expectation that quantum devices may outperform classical ones on certain tasks~\cite{preskill2012quantum}. While specific quantum algorithms have demonstrated exponential or quadratic speed-ups, e.g. Shor~\cite{shor1994algorithms} and Grover~\cite{grover1996fast}, the empirical evidence supporting a potential quantum advantage in the context of quantum machine learning (QML), including quantum generative modelling, remains mixed~\cite{schuld2022quantum,bowles2024better,rudolph2024trainability}.
In this work, we focus on hybrid Quantum Generative Adversarial Networks (QuGANs), that integrate parametrized quantum circuits into the adversarial training framework for generating new data. Of particular interest is the generator, which, once trained, can serve as a quantum-native mechanism for loading data distributions into quantum states, potentially mitigating the input bottleneck that threatens quantum speed-up~\cite{zoufal2019quantum}.
While prior work has largely centred around generic, hardware-efficient quantum circuit architectures, we investigate how domain-specific structure, in particular geometric priors, can be embedded into the quantum circuit design via different entanglement-schemes to improve generative performance of the quantum generator. We complement our circuit-design search with two model-agnostic enhancements: a variance regularization term to foster output-variety, and a classical output-scaling layer for improved distributional control. Our task is to generate fully connected weighted graphs ($K_4$), where edge weights represent distances in three-dimensional Euclidean space. This setting is not only representative of real-world applications, but also imposes a challenging structural constraint: the edge weights must jointly satisfy the triangle inequality (for every triplet of nodes) and the Ptolemaic inequality (across each quadrilateral). While existing literature on QuGANs has mostly evaluated models based on distributional fidelity, thus overlooking the internal dependencies and constraints that characterize structured data, we also take into account these complex dependencies between edge weights. To evaluate both statistical and geometric aspects of generative performance, we introduce the Triangle Validity Score (TVS) and the Four-Point Ptolemaic Consistency Metric (4PCM). Our experiments reveal that problem-aligned circuit designs can significantly improve geometric validity while preserving strong distributional fidelity. Moreover, we identify clear trade-offs between fidelity and constraint satisfaction, governed by fine-grained design choices.

\section{BACKGROUND}
\label{sec:background}
\subsection{Graph Representation and Geometric Validity} \label{sec:graph_representation}
We consider the complete weighted graph, denoted as $K_n$, which consists of $n$ vertices and $d = \frac{n(n-1)}{2}$ edges connecting every distinct pair of vertices. The graph can be expressed uniquely by a $d$-dimensional vector $\mathbf{e}$, where each component $e_{ij}$ represents the weight of the edge between vertices $i$ and $j$. This formulation is well-suited for modelling various real-world systems, such as spatial networks where vertices represent locations and edge weights correspond to the distances between them. \\
When the edge weights are normalized to sum to one, the vector $\mathbf{e}$ lies in the $(d-1)$-dimensional simplex:
\[
\Delta^{d-1} = \left\{ \mathbf{e} \in \mathbb{R}^d \,\middle|\, e_i \geq 0 \ \text{for all } i,\quad \sum_{i=1}^{d} e_i = 1 \right\}
\]
However, only a subset of this simplex corresponds to geometrically valid graphs, as the edge weights must also satisfy additional constraints arising from Euclidean geometry.
A graph is considered \textit{geometrically valid} if its edge weights, interpreted as distances, satisfy two key metric constraints: the \textit{Triangle Inequality} for every three-node subset and the \textit{Ptolemaic Inequality} for every four-node quadrilateral. The Triangle Inequality states that for any two edge lengths, their sum must be greater than or equal to the third. For any three vertices $\{i, j, k\}$, this can be expressed as:
\[
e_{ij} + e_{jk} \geq e_{ik}
\]
The Ptolemaic Inequality requires that the product of the lengths of a quadrilateral's two diagonals be less than or equal to the sum of the products of its opposite side pairs. For any four node quadrilateral $ijkl$, this can be formulated as:
\[
e_{ik} \cdot e_{jl} \leq e_{ij} \cdot e_{kl} + e_{jk} \cdot e_{il}
\]
The triangle inequality ensures local three-point consistency, while the Ptolemaic inequality governs the global relationship in Euclidean space. Violations of these constraints lead to unrealistic structures with distorted edge lengths and invalid physical embeddings, thereby compromising the network's integrity and interpretability.

\subsection{Generative Adversarial Networks} \label{sec:gan}
The Generative Adversarial Network (GAN) paradigm formulates the problem of generative learning as a zero-sum game between two competing neural networks: a generator \(G\) and a discriminator \(D\). The generator, \(G_{\boldsymbol{\theta}}\), learns to map random noise vectors \(\mathbf{z}\) from a latent distribution \(p_z\) to synthetic data samples \(\mathbf{e'} = G_{\boldsymbol{\theta}}(\mathbf{z})\). Its objective is to produce samples that are indistinguishable from those of the real data distribution, \(\mathcal{E}_{\text{real}}\). Concurrently, the discriminator, \(D_{\boldsymbol{\phi}}\), is trained to distinguish these generated samples from genuine data. It outputs a probability that a given sample is real. The two networks are trained in an adversarial process: \(G\) is optimized to produce samples that \(D\) misclassified as real, while \(D\) is optimized to correctly identify forgeries. This dynamic is captured by the standard minimax objective, where \(G\) seeks to minimize the value function that \(D\) attempts to maximize:

\begin{align*}
\min_{\boldsymbol{\theta}} \; \max_{\boldsymbol{\phi}} \;
& \mathbb{E}_{\mathbf{e} \sim \mathcal{E}_{\text{real}}}
\left[\log D_{\boldsymbol{\phi}}(\mathbf{e})\right] \\
+ & \mathbb{E}_{\mathbf{z} \sim p_z}
\left[\log\left(1 -
D_{\boldsymbol{\phi}}\left(G_{\boldsymbol{\theta}}(\mathbf{z})\right)
\right)\right]
\end{align*}

At equilibrium, the generator captures the target data distribution, and the discriminator's output for any sample is \(\frac{1}{2}\), indicating it cannot distinguish real from fake samples better than random chance. \\
While GANs have emerged in the field of classical (non-quantum) AI, implementations using QC have also been developed. A distinction is made between hybrid QuGANs, where either \(G\) or \(D\) is realized by means of a trainable parametrized quantum circuit (PQC) and the counterpart is a classical neural network, or, fully QuGANs, where both adversaries are realized through a PQC.

\section{RELATED WORK}
\label{sec:relatedwork}
\subsection{Generative Models for Graph-Structured Data.} 
Generative models for graph-structured data can broadly be categorized into two paradigms: sequential generation and one-shot generation~\cite{zhu2022graphsurvey}. Sequential generation models construct graphs incrementally, adding nodes or edges over multiple steps~\cite{you2018graphrnn}. This approach enables intermediate constraint checking, often resulting in higher validity of generated graphs with respect to their imposed constraints~\cite{Shi2020GraphAF}. In contrast, one-shot generation models generate entire graphs in a single pass, typically requiring post-processing to enforce constraints due to their limited ability to incorporate validity checks during generation~\cite{zang2020moflow}. GANs typically follow the one-shot strategy. A fundamental research gap of graph generation frameworks with regard to evaluation metrics is the primary focus on statistics-based metrics (e.g. the reliance on distributions), while self-quality-based metrics like validity tend to be under-represented~\cite{zhu2022graphsurvey}.

\subsection{Quantum Generative Adversarial Networks.} 
Based on the developments of classical GANs, QuGANs were independently proposed by Dallaire–Demers \& Killoran~\cite{dallaire2018quantum} and by Lloyd \& Weedbrook~\cite{lloyd2018quantum} as a concept for learning probability distributions utilizing QC. Numerous specialisations have followed, including Quantum Wasserstein GANs~\cite{chakrabarti2019qwgan} and Tensor-Network QuGANs~\cite{huggins2019tnqgan}. Empirically speaking, QuGANs with a comparable or even smaller number of parameters have matched or outperformed classical GANs in areas such as image synthesis~\cite{stein2021qugan} or generative chemistry~\cite{kao2023qganchem}, with the models demonstrating stable performance even under moderate NISQ noise~\cite{anand2021noisequgan}. Among existing works, Rohe \textit{et al.}~\cite{rohe2025searoute} are closest to ours, introducing self-quality-based metrics via triangle inequality checks on sea-route graphs; however, we extend this approach by incorporating more comprehensive geometric constraints and by systematically evaluating a broader portfolio of quantum architectures.

\subsection{Quantum Circuit Ansätze.} The performance of variational quantum algorithms hinges on their circuit, also called \emph{Ansatz}~\cite{sim2019Expressibility}, which dictates the model's inductive biases. A key distinction exists between hardware-efficient layouts, which prioritize shallow, device-native gates~\cite{kandala2017hardware}, and problem-inspired circuits, which embed, for example, domain symmetries directly into the entanglement pattern~\cite{meyer2023exploiting,bako2024problem}. It remains an open question whether a problem topology-aligned quantum circuit can learn the underlying data distribution (referring to statistics-based metrics) more effectively than a classical GAN, while at the same time inherently learning and considering the validity of graphs (referring to self-quality-based metrics) better.

\section{METHODOLOGY}
\label{sec:methodology}
\subsection{(Qu)GAN Architectures}
The classical baseline generator with $84$ parameters is a single–hidden-layer multilayer perceptron (MLP) that maps a 6D Gaussian latent vector to a 6D edge–weight vector via a 6-unit hidden layer and a \texttt{LeakyReLU} activation function, followed by a linear layer and a \texttt{Softmax} activation. The discriminator is a three hidden–layer MLP applied uniformly across all models (classical and quantum), ensuring that performance differences arise from the generator class rather than the adversary.
Combining our focus on the generation of artificial graphs with robust comparability, we have opted for hybrid QuGANs. In our case, the classical generator is replaced by a PQC operating on a 6-qubit register, with one qubit assigned to each edge of the $K_4$ graph using a fixed edge–to–wire mapping. Assuming the quantum register is initialized in the state $\ket{0}^{\otimes d}$, each PQC follows a common four-stage pipeline: (i) encode a 6D latent vector $z \sim \mathcal{N}(0,I)$ into single-qubit rotations; (ii) apply a variational, trainable unitary $U(\boldsymbol{\theta})$ whose entanglement pattern defines the model variant; (iii) measure the Pauli-Z expectation value $\langle \sigma^z_i \rangle \in [-1,1]$; (iv) affinely map to $[0,1]$ and renormalize to the probability simplex to yield an edge-weight vector consumable by the fixed classical discriminator. Apart from entanglement layout (and associated gate choices), all training hyperparameters are held constant across variants, and each QuGAN consists of $5$ layers with a total of $90$ trainable parameters.

\begin{figure*}[!t]
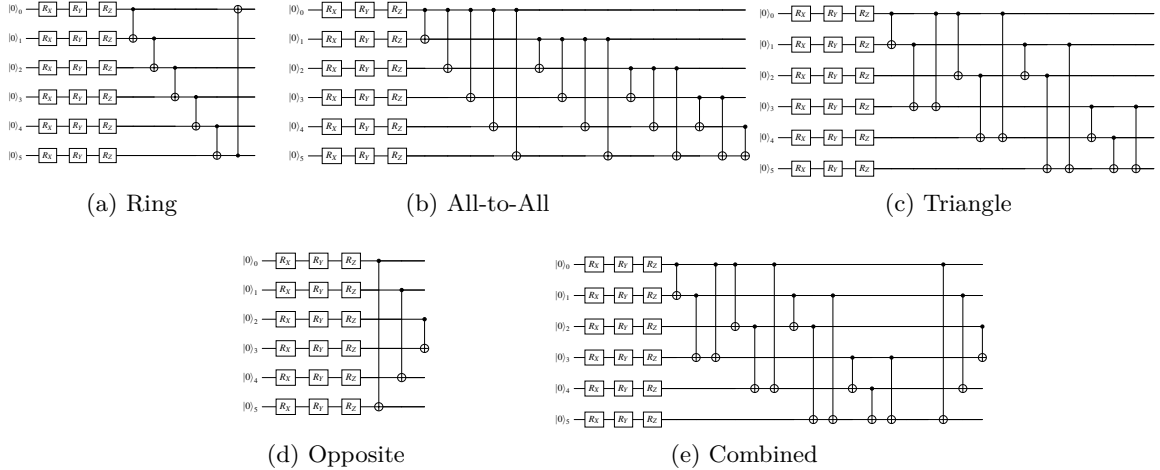

\centering

\newlength{\circuitpanelheight}
\setlength{\circuitpanelheight}{1.2cm}  

\newcommand{\circuitpanel}[1]{%
  \parbox[t][\circuitpanelheight][c]{\linewidth}{%
    \centering
    \resizebox{!}{\circuitpanelheight}{\input{#1}}%
  }\\[1.5em]  
}

\begin{subfigure}[t]{0.32\textwidth}
  \centering
  \parbox[t][\circuitpanelheight][c]{\linewidth}{%
    \centering
    \resizebox{!}{\circuitpanelheight}{\input{tikz/ring}}%
  }\\[1.5em]  

  \caption{Ring}
  \label{fig:ent_ring}
\end{subfigure}\hfill
\hspace*{-5em} 
\begin{subfigure}[t]{0.32\textwidth}
  \centering
  \parbox[t][\circuitpanelheight][c]{\linewidth}{%
    \centering
    \resizebox{!}{\circuitpanelheight}{\input{tikz/all}}%
  }\\[1.5em]  

  \caption{All-to-All}
  \label{fig:ent_all}
\end{subfigure}\hfill
\begin{subfigure}[t]{0.32\textwidth}
  \centering
  \parbox[t][\circuitpanelheight][c]{\linewidth}{%
    \centering
    \resizebox{!}{\circuitpanelheight}{\input{tikz/structure}}%
  }\\[1.5em]  
  \caption{Triangle}
  \label{fig:ent_triangle}
\end{subfigure}

\vspace{3.25em}

\makebox[0.66\textwidth]{
  \begin{subfigure}[t]{0.32\textwidth}
    \centering
  \parbox[t][\circuitpanelheight][c]{\linewidth}{%
    \centering
    \resizebox{!}{\circuitpanelheight}{\input{tikz/opposite}}%
  }\\[1.5em]  

    \caption{Opposite}
    \label{fig:ent_opposite}
  \end{subfigure}\hfill
  \begin{subfigure}[t]{0.32\textwidth}
    \centering
  \parbox[t][\circuitpanelheight][c]{\linewidth}{%
    \centering
    \resizebox{!}{\circuitpanelheight}{\input{tikz/combined}}%
  }\\[1.5em]  

    \caption{Combined}
    \label{fig:ent_combined}
  \end{subfigure}%
}
\begin{center}
    \vspace{0.3cm}
    \caption{Generator circuit topologies evaluated in this work. (a) Ring, (b) All-to-All, (c) Triangle, (d) Opposite, and (e) Combined. Each panel depicts a single, representative variational layer (single-qubit rotations followed by a specified two-qubit entangling topology).}
    \label{fig:entanglement_strategies}
\end{center}
\end{figure*}

We evaluate five PQC generator topologies, ordered from generic to problem-informed:
\begin{itemize}
  \item \textbf{Ring:} Local, translation-style inductive bias; two-qubit gates couple qubits in a closed nearest-neighbor loop.
  \item \textbf{All-to-All:} Maximally connected reference with entangling gates between every qubit pair; capacity-oriented baseline.
  \item \textbf{Triangle:} Geometry-aligned design; qubits corresponding to the three edges of each of the four triangular subgraphs of $K_4$ are fully entangled, embedding triangle structure as a circuit prior.
  \item \textbf{Opposite:} Data-driven sparse layout; entangles only qubit pairs representing empirically anti-correlated (opposite) edges observed in the training data.
  \item \textbf{Combined:} Union of Triangle and Opposite couplings to encode both geometric structure and learned global correlations.
\end{itemize}
All fundamental QuGAN architectures are illustrated in Fig.~\ref{fig:entanglement_strategies}.

\subsection{Triangle Topology Variants}
To better understand the role of entanglement in our best performing quantum generator (QuGAN~Triangle), we study it in greater detail by introducing variations of its two-qubit interaction layer. Our goal is to isolate which aspects of the entanglement scheme drive performance differences.

\begin{figure}[t]
\centering
\scalebox{0.45}{\input{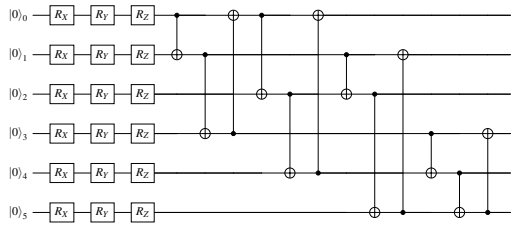}}
\begin{center}
    \vspace{0.3cm}
    \caption{The \textsc{Triangle Cyclic CNOT} entanglement scheme.  The \textsc{Triangle Non-Cyclic CNOT} variant is shown in Figure~\ref{fig:ent_triangle}. \textsc{Triangle Non-Cyclic CRot} and \textsc{Triangle Cyclic CRot} follow the same layout as their CNOT counterparts but replace the CNOT gates with trainable controlled-$R_Y(\theta)$ gates.}
    \label{fig:triangle_topologies}
\end{center}
\end{figure}

We define a $2\times2$ family of \emph{triangle-inspired} architectures built on the same encoding and single-qubit rotation layers as the base Triangle topology; only the entangling layer is slightly modified (Fig.~\ref{fig:triangle_topologies}). Variation occurs along two design axes:

\begin{enumerate}
  \item \textbf{Entanglement ordering (directionality).} We compare a standard, non-cyclic application (control wires determined by index; no directional bias) against a directed \emph{cyclic} sequence that may lead to an ``ordered flow'' of entanglement and thus a stronger inductive bias.
  \item \textbf{Gate type (interaction strength control).} We contrast fixed CNOT gates with trainable controlled-$R_Y(\theta)$ rotations, permitting fine-grained adjustment of entanglement.
\end{enumerate}

Crossing these axes yields us four variants — (Non-Cyclic/Cyclic) \(\times\) (CNOT/CRot). To ensure a fair comparison with the same number of parameters, all CRot variants have only 3 layers, which also gives them exactly $90$ parameters.

\subsection{Model Enhancements}
\subsubsection{Variance Control.} Across all QuGAN variants we observed a persistent \emph{underestimation of variance} relative to the empirical edge‑weight distribution: generated batches tended to concentrate too narrowly in the simplex. This raises an attribution question: is the reduced variance a fundamental limitation of the circuit expressivity itself, or a consequence of the adversarial training signal (e.g., the discriminator not penalizing under‑dispersion strongly, or an objective mismatch in the generator loss). To probe this, we augment the generators loss function with an optional \emph{variance matching} term that explicitly penalizes deviations from the distributional spread of the target data.

For the training batch $\left\{ e^{(b)} \right\}_{b=1}^{B}$, with $\quad e^{(b)} \in \mathbb{R}^d$, the batch standard deviation is defined as 
\[
\sigma_{\text{batch}} = \frac{1}{B} \sum_{b=1}^{B} \text{Std}_d\left(e^{(b)}\right).
\]
Analogously $\sigma_{\mathrm{target}}$, the standard deviation of the real data $\left\{ x^{(n)} \right\}_{n=1}^{N}$, is calculated. We then define
\[
L_{\mathrm{variance}}
= \bigl(\sigma_{\mathrm{batch}} - \sigma_{\mathrm{target}}\bigr)^{2}.
\label{eq:variance_loss}
\]
This scalar penalty is weighted by a hyperparameter $\lambda$ and added to the primary adversarial generator loss $L_{G,\mathrm{BCE}}$, yielding
\[
L_{G,\mathrm{total}} = L_{G,\mathrm{BCE}} + \lambda \, L_{\mathrm{variance}}.
\label{eq:total_loss}
\]

For consistency, the variance regularization is evaluated solely on the QuGAN Triangle architecture, resulting in the variations \textsc{Triangle~Loss} and \textsc{Triangle~Loss+Scale}.

\subsubsection{Output Scaling.} 
To test whether a lightweight classical scaling can improve marginal fit without altering the circuit’s learned correlational structure, we introduce an optional \emph{output‑scaling layer} applied in post‑processing to the generator outputs.

Let $\mathbf{v} \in [-1,1]^d$ denote the vector of expectation values obtained from $Z$‑basis measurements of the $d$ qubits. The transformation proceeds in three deterministic steps:

1. Affine shift to non-negative range
\[
\mathbf{v}' \gets \frac{1 + \mathbf{v}}{2}, \qquad \mathbf{v}' \in [0,1]^d .
\]

2. Per‑channel rescaling
with learnable non-negative scale factors $\boldsymbol{\alpha} \in \mathbb{R}_+^d$,
\[
\mathbf{s} \gets \boldsymbol{\alpha} \odot \mathbf{v}',
\]
where $\odot$ denotes elementwise multiplication. This step adjusts the \emph{relative} amplitudes across edge dimensions.

3. Projection to the simplex
\[
\mathbf{w} \gets \frac{\mathbf{s}}{\sum_i s_i + \varepsilon}, \qquad \varepsilon > 0 \text{ (stability)},
\]
yielding normalized edge weights $\mathbf{w} \in [0,1]^d$, $\sum_i w_i = 1$.

Because the final normalization removes any common multiplicative factor, the scaling layer can only reweight edges relative to one another. Once fixed, the scaling layer acts as a lightweight calibration tool, while the PQC remains responsible for capturing the core dependency structure. We evaluate this enhancement in conjunction with the variance matching term on the QuGAN~Triangle model (reported as \textsc{QuGAN~Triangle~Loss+Scale}).

\subsection{Experimental Setup}
\subsubsection{Dataset Generation \& Preprocessing.} We construct a dataset of normalized four‑airport distance sextuples from the OpenFlights global airport database~\cite{openflights2017}, which is representative of many graph structures in application. After filtering for airports, \(7{,}698\) remained. We then uniformly sample \(10{,}000\) unique unordered quadruples; each defines the vertex set of a complete \(K_4\) graph. The problem size was chosen to ensure classic simulability, although this is not yet a size for a real use case. For every quadruple, the six pairwise geodesic distances are computed via the haversine formula~\cite{sinnott1984haversine}, yielding edge lengths grounded in realistic spherical geometry.
To avoid degenerate cases with extremely short links—which produce numerically tiny normalized weights and destabilize training—we discard any \(K_4\) containing an edge shorter than \(185.2\,\mathrm{km}\) (\(\approx 100\)~nautical miles). Let \(\mathbf{e}=(e_1,\dots,e_6)\) denote the edge lengths, we scale each sextuple to the unit‑sum constraint
\[
\mathbf{e}' \;=\; \frac{\mathbf{e}}{\sum_{i=1}^{6} e_i},
\]
so that \(\mathbf{e}' \in \Delta_5\) (the 5‑simplex). This normalization renders the learning task scale‑invariant with respect to absolute distance while preserving the relative geometric structure encoded in the edge ratios.

\subsubsection{Training and Hyperparameters.} All GAN architectures were trained for 1,000~epochs using \textsc{PyTorch}~\cite{paszke2019pytorch}, with quantum circuits implemented in \textsc{PennyLane}~\cite{bergholm2018pennylane} on the \texttt{default.qubit} simulator random seeded.  
To ensure a strictly fair comparison, all models were trained with the same hyperparameter settings, summarised in Tab.~\ref{tab:hyperparams}. Unless noted otherwise, these settings are kept fixed across architectures so that performance differences can be attributed solely to circuit topology and gate choice.

\begin{table}[t]
\centering
\small   
\begin{tabularx}{\columnwidth}{@{}l X@{}}
\toprule
\textbf{Hyperparameter} & \textbf{Value / Setting} \\
\midrule
Optimizer & Adam ($\beta_1 = 0.5$, $\beta_2 = 0.999$) \\
Learning rates & $\eta_G = 5\times10^{-4}$; $\eta_D = 2\times10^{-4}$ \\
Batch size & 64 \\
Label smoothing & Real labels shifted to 0.9 \\
Input noise & 6D i.i.d.\ $\mathcal{N}(0,1)$ \\
QNode gradient & Parameter-shift rule \\
PQC depth & 5 layers (CNOT); 3 layers (CRY) \\
Entangling init & $CR_Y(\theta_{ij})$, $\theta_{ij} \sim \mathcal{N}(0,0.01)$ \\
Variance regularizer & Enabled ($\lambda = 5000$) \\
Geometry tolerances & Triangle: $10^{-2}$; Ptolemy: $10^{-2}$ \\
Evaluation size & 5{,}000 samples/model \\
\bottomrule
\end{tabularx}
\caption{Unified hyperparameter configuration (applies to all models unless noted).}
\label{tab:hyperparams}
\end{table}

\subsubsection{Evaluation Metrics} Our evaluation metrics target two aspects: \emph{statistical fidelity} with regards to the empirical edge weight distribution and \emph{geometric validity} under Euclidean constraints.

\begin{description}[leftmargin=0pt,labelsep=0.5em,font=\normalfont\itshape]
\item[Statistical fidelity]
We report two measures: (i) the 1‑Wasserstein distance \(W\) between empirical and generated \emph{marginal} edge‑weight distributions (lower is better)~\cite{villani2009optimal}; and (ii) the Jensen-Shannon divergence (JS) between discretized empirical and generated distributions (lower is better)~\cite{lin1991divergence}.

\item[Geometric validity]
A generated \(K_4\) graph with its six edge weights is considered \emph{triangle‑valid} if for every 3-node subset of the graph the triangle inequality holds with tolerance of \(\varepsilon = 10^{-2}\). The \emph{Triangle Validity Score} (TVS) is the fraction of samples satisfying all four such tests.
Global four‑point consistency is assessed by the \emph{Four‑Point Ptolemaic Consistency Metric} (4PCM). For each generated \(K_4\) graph, all three perfect matchings of opposite edge pairs are checked; a sample passes if each satisfies the Ptolemaic ineqaulity. 4PCM is the fraction of samples passing all three inequalities. Both TVS and 4PCM are reported as percentages (higher is better).
\end{description}

\section{EXPERIMENTAL RESULTS \& ANALYSIS}
\label{sec:results}
\begin{table*}[ht]
\centering
\renewcommand{\arraystretch}{1.2}
\setlength{\tabcolsep}{10pt}
\resizebox{\textwidth}{!}{%
\begin{tabular}{lcccccc}
\toprule
\textbf{Model Name} & \textbf{Wass.} & \textbf{JS Div.} & \textbf{TVS (\%)} & \textbf{TVS (95\% CI)} & \textbf{4PCM (\%)} & \textbf{4PCM (95\% CI)} \\
\midrule
\multicolumn{7}{l}{\textbf{Architectural Comparison}} \\
GAN Classic                    & \textbf{0.008} [0.008,\ 0.008] & \textbf{0.117} [0.117,\ 0.117] & 69.14 & [63.89,\ 74.40] & \textbf{91.28} & [89.84,\ 92.72] \\
QuGAN Ring                     & 0.018 [0.018,\ 0.018] & 0.237 [0.237,\ 0.237] & 76.08 & [71.53,\ 80.62] & 89.36 & [86.24,\ 92.49] \\
QuGAN All-to-All               & 0.017 [0.017,\ 0.017] & 0.239 [0.239,\ 0.239] & 75.50 & [73.94,\ 77.05] & 88.94 & [87.34,\ 90.54] \\
QuGAN Triangle                 & 0.015 [0.015,\ 0.015] & 0.197 [0.197,\ 0.197] & \textbf{78.74} & [74.01,\ 83.48] & 90.90 & [86.77,\ 95.03] \\
QuGAN Opposite                 & 0.020 [0.020,\ 0.020] & 0.283 [0.283,\ 0.283] & 70.57 & [66.39,\ 74.75] & 88.20 & [85.44,\ 90.96] \\
QuGAN Combined                 & 0.018 [0.018,\ 0.018] & 0.255 [0.255,\ 0.255] & 77.44 & [75.75,\ 79.12] & 89.23 & [87.12,\ 91.34] \\
\addlinespace[0.5em]
\multicolumn{7}{l}{\textbf{Triangle Topology Variants}} \\
QuGAN Triangle Cyclic CNOT     & 0.020 [0.018,\ 0.021] & 0.268 [0.247,\ 0.290] & 76.91 & [75.83,\ 78.04] & \textbf{89.69} & [87.74,\ 91.65] \\
QuGAN Triangle Cyclic CRot     & 0.015 [0.013,\ 0.016] & 0.205 [0.193,\ 0.216] & 75.36 & [71.11,\ 79.60] & 89.02 & [87.44,\ 90.61] \\
QuGAN Triangle Non-Cyclic CNOT & 0.016 [0.013,\ 0.018] & 0.226 [0.202,\ 0.250] & \textbf{78.83} & [75.83,\ 81.84] & \textbf{89.69} & [87.74,\ 91.64] \\
QuGAN Triangle Non-Cyclic CRot & \textbf{0.014} [0.012,\ 0.016] & \textbf{0.200} [0.167,\ 0.234] & 75.26 & [72.03,\ 80.49] & 89.32 & [87.45,\ 91.19] \\
\addlinespace[0.5em]
\multicolumn{7}{l}{\textbf{Model Enhancements}} \\
QuGAN Triangle Loss            & 0.009 [0.009,\ 0.009] & 0.137 [0.137,\ 0.137] & 73.42 & [65.15,\ 81.68] & 89.87 & [83.80,\ 95.94] \\
QuGAN Triangle Loss Scale      & \textbf{0.008} [0.008,\ 0.008] & 0.136 [0.136,\ 0.136] & 75.72 & [70.52,\ 80.92] & 90.89 & [88.48,\ 93.31] \\
\bottomrule
\end{tabular}%
}
\caption{Evaluation metrics for all generator architectures and variants.  
Lower values are better for distribution metrics; higher is better for validity scores.  
For ``Architectural Comparison'' and ``Triangle Topology Variants'' best results were printed bold.}
\label{tab:final_results_combined}
\end{table*}

\subsection{Architectural Comparison}
\begin{figure}[t]
  \centering
  \includegraphics[width=1\linewidth]{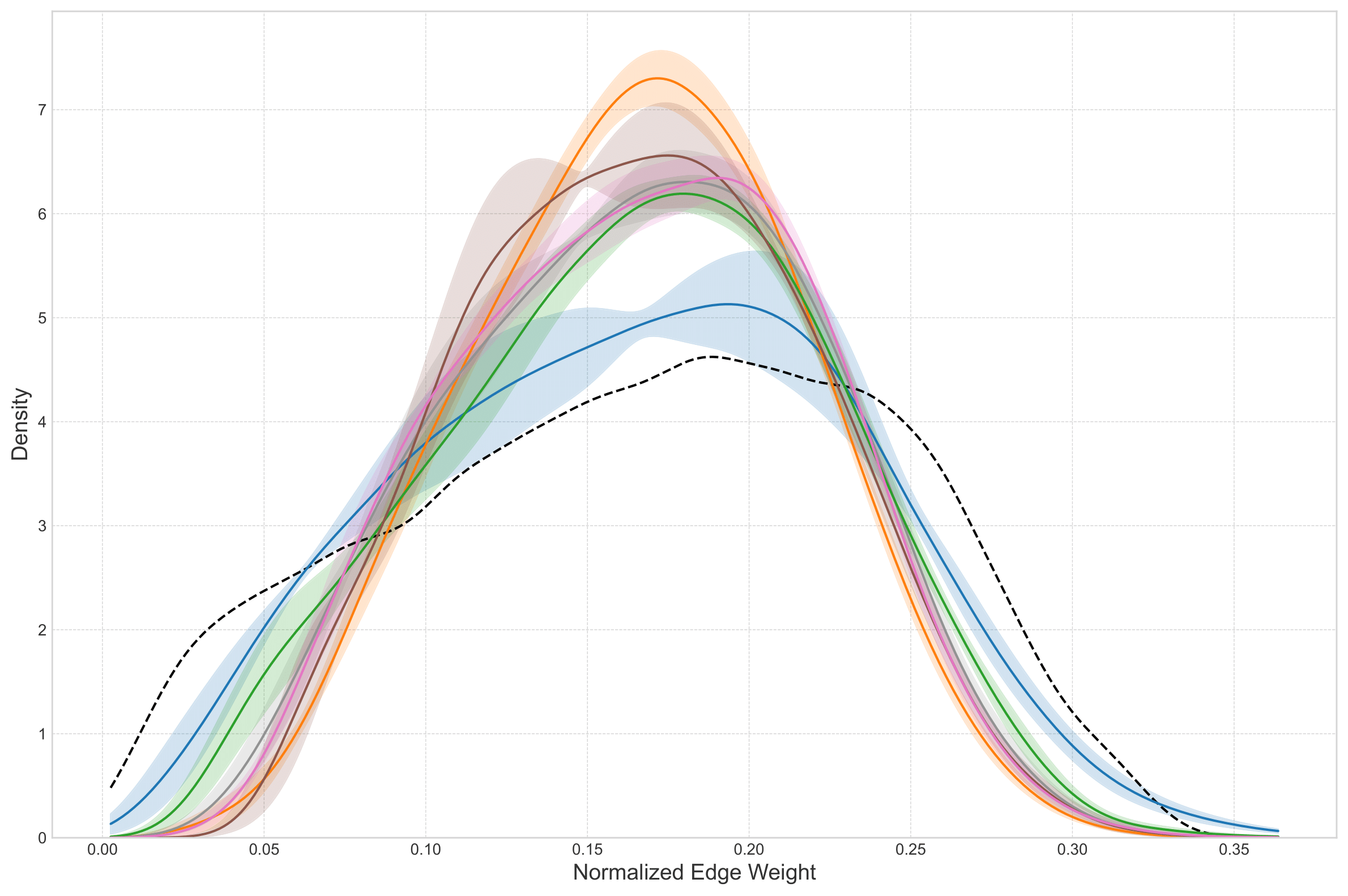}
  

  \includegraphics[width=0.8\linewidth]{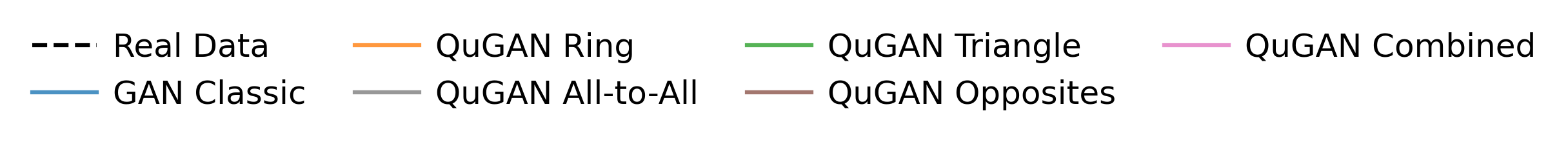}
  
  \caption{
  Learned edge weight distributions for each generator architecture.}
  \label{fig:edge_distributions}
\end{figure}

Our architectural comparison across generator architectures (Fig.~\ref{fig:edge_distributions}, Tab.~\ref{tab:final_results_combined}) reveals distinct inductive biases. With regards to the edge-weight distributions, the classical GAN baseline achieves the closest match to the target distribution, exhibiting a broad spread, moderate skewness, and a relatively low peak amplitude. Among the quantum models, the \textsc{Triangle} topology yields the best approximation, even if it does not come as close as the classical GAN. In contrast, the \textsc{Ring} topology produces distributions that most closely resemble a standard normal distribution.
Interestingly, both the \textsc{Opposite} topology and the classical GAN display signs of bimodality, as reflected by the shaded variance bands. While the real distribution is not explicitly bimodal, this behaviour may implicitly try to capture the left-skewed structure.
The strong performance of the \textsc{Triangle} topology and the distinctive distribution of the \textsc{Opposite} model suggest that their constrained, task-specific designs, arguably the least generic among all topologies, may confer a meaningful advantage. This is especially plausible in the case of the Opposite architecture, which embeds statistical properties of the training data directly into its entanglement pattern.

Quantitatively, \textsc{QuGAN~Triangle} achieves the strongest overall performance among all QuGAN architectures, obtaining the lowest Wasserstein ($0.015$) and JS divergence ($0.197$) values in that group, together with the highest geometric scores (TVS \(=78.74\%\), 4PCM \(=90.90\%\)). When compared to the \textsc{GAN~Classic} baseline, trade‑offs emerge: the classical model attains superior distributional fidelity (Wass.\ $0.008$; JS $0.117$) and slightly higher 4PCM ($91.28\%$), whereas \textsc{QuGAN~Triangle} substantially improves local geometric consistency (TVS \(78.74\%\) vs.\ \(69.14\%\)), a strength we have expected through the specific circuit design. The 4PCM gap between the two is modest given the reported confidence intervals.

Given this balance of strengths, we select the \textsc{QuGAN~Triangle} architecture as the reference quantum model for deeper analysis.

\subsection{Triangle Topology Variants}
Results in Tab.~\ref{tab:final_results_combined} indicate that \emph{gate type} is a relevant factor: CNOT‑based variants yield stronger geometric validity, whereas controlled‑$R_Y$ (CRot) variants improve distributional fidelity.

The \textsc{QuGAN~Triangle~Non\mbox{-}Cyclic~CNOT} attains the highest TVS (\(78.83\%\)) and shares the top 4PCM value (\(89.69\%\)) within the family; its cyclic CNOT counterpart is close (TVS \(76.91\%\), 4PCM \(89.69\%\)). Both CNOT models exhibit higher Wasserstein / JS divergence (up to \(0.020\) / \(0.268\)) than their CRot analogues. In contrast, \textsc{QuGAN~Triangle~Non\mbox{-}Cyclic~CRot} achieves the best statistical fit (Wass.\ \(0.014\), JS \(0.200\)), with the cyclic CRot variant close behind (Wass.\ \(0.015\), JS \(0.205\)); these gains come with modest reductions in geometric scores (TVS \(\approx 75\%\), 4PCM \(\approx89\%\)). The differences between cyclic and non-cyclic arrangements are uniformly smaller than the differences due to gate type, suggesting that the parametrization of the entanglement interaction is the primary trade-off between geometric consistency and distributional accuracy in this architecture. The exact training performance can be found in the supplementary material.

\subsection{Model Enhancements}
\begin{figure*}[t]
  \centering
  \begin{minipage}[t]{0.45\textwidth}
    \centering
    \includegraphics[width=\linewidth]{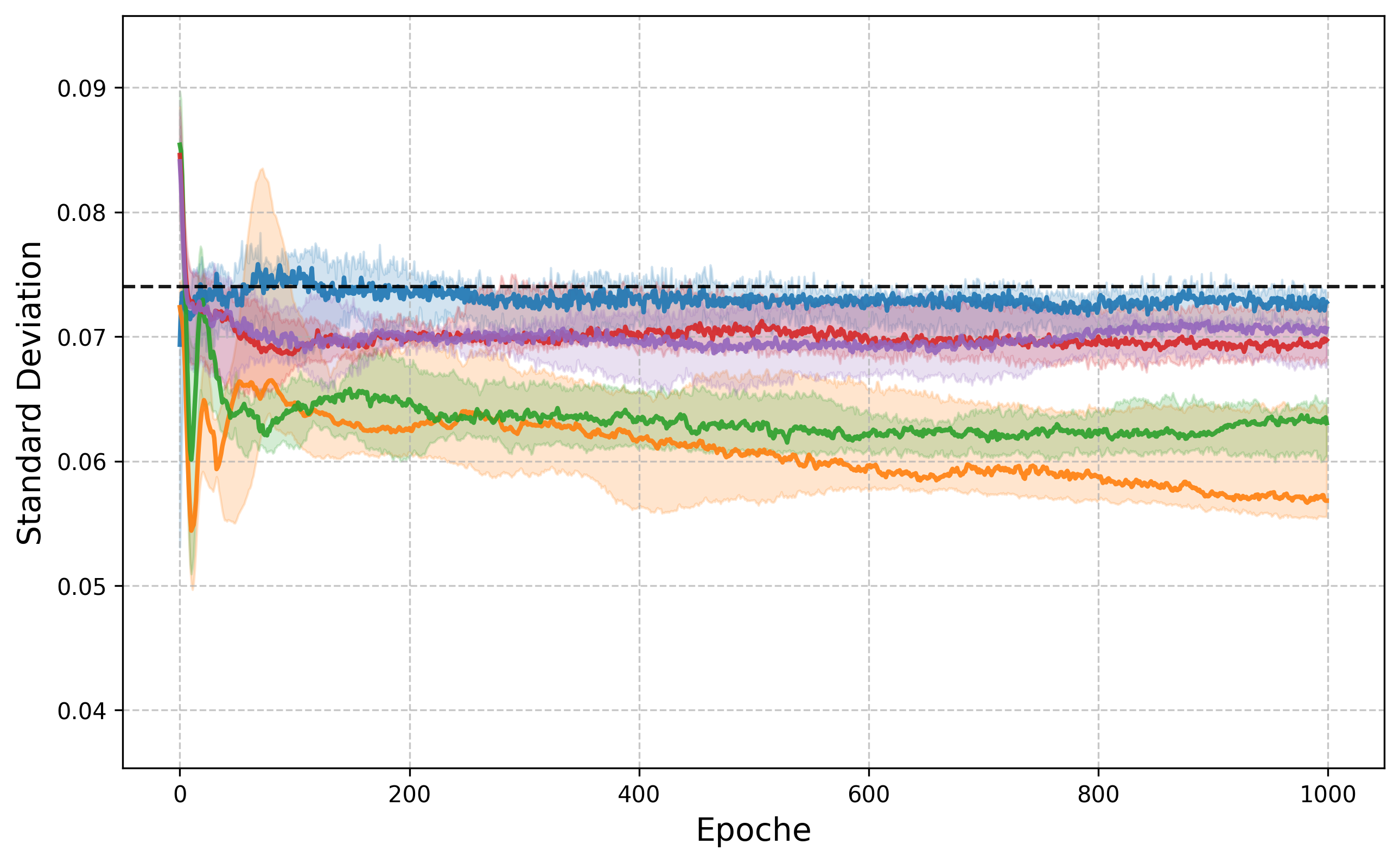}
    \caption*{(a) Standard Deviation over Epochs}
  \end{minipage}
  \hfill
  \begin{minipage}[t]{0.45\textwidth}
    \centering
    \includegraphics[width=\linewidth]{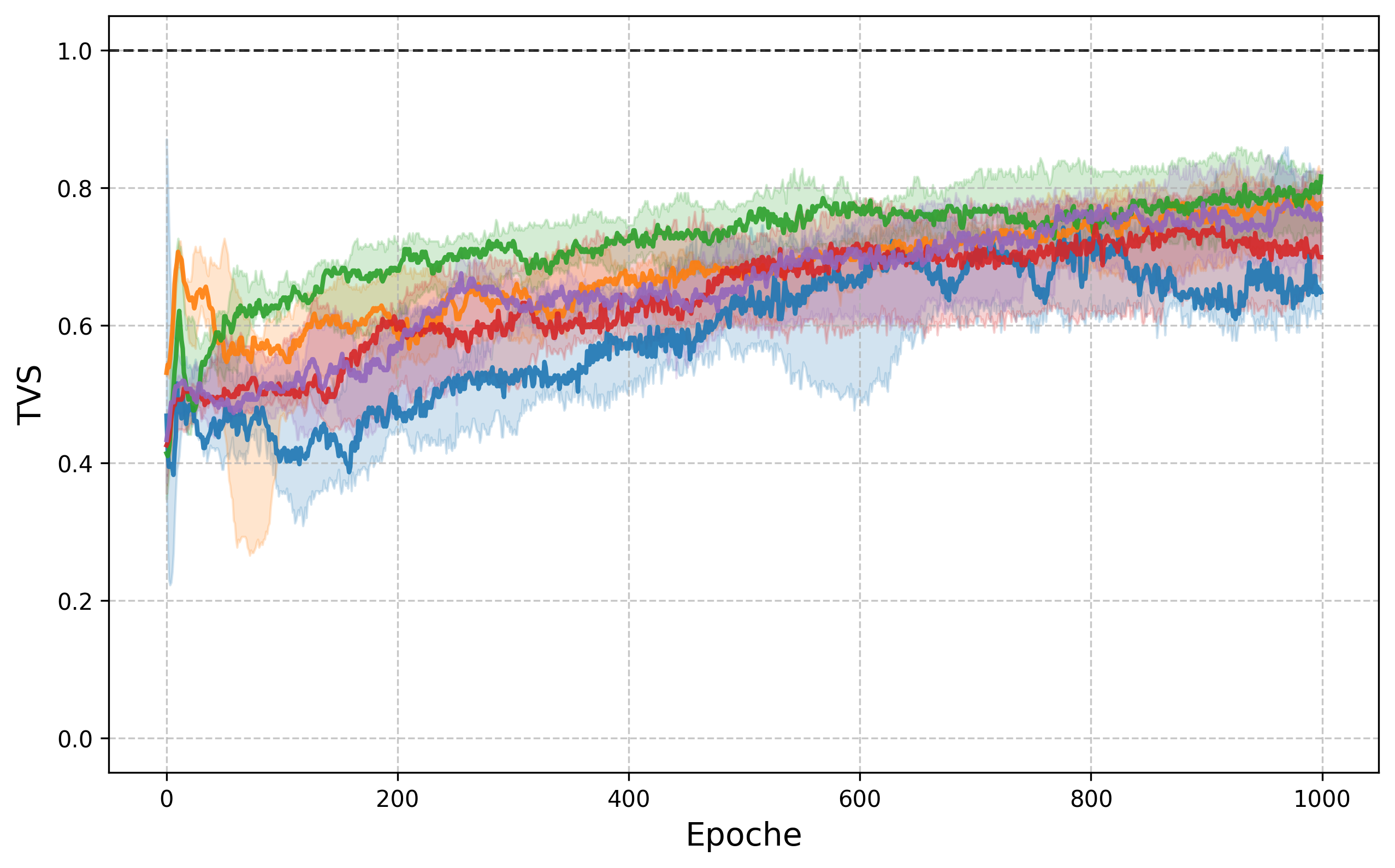}
    \caption*{(b) Triangle Validity over Epochs}
  \end{minipage}
  \vspace{0.5cm}

  \begin{minipage}[t]{0.45\textwidth}
    \centering
    \includegraphics[width=\linewidth]{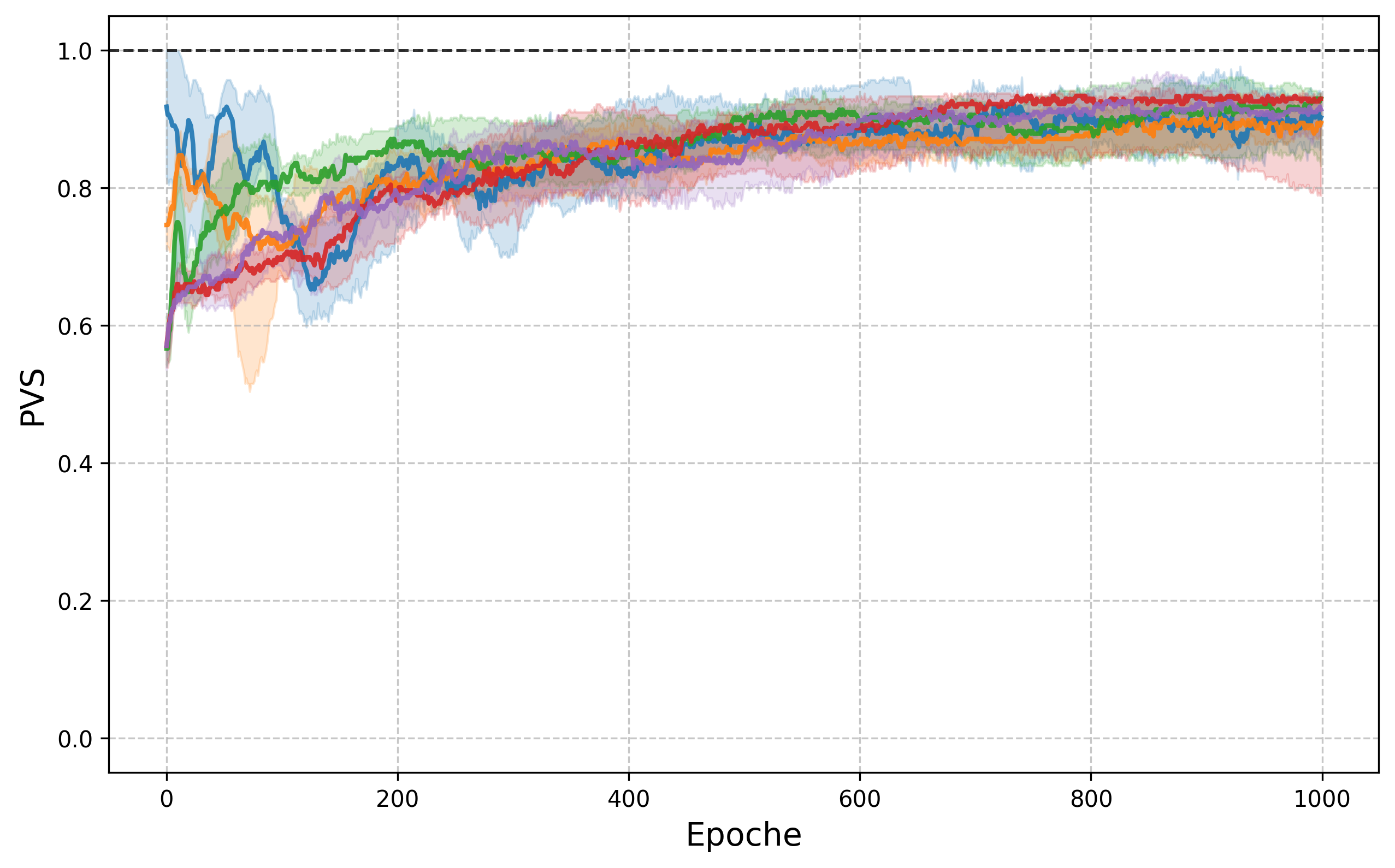}
    \caption*{(c) Ptolemaic Validity over Epochs}
  \end{minipage}
  \hfill
  \begin{minipage}[t]{0.45\textwidth}
    \centering
    \raisebox{0.075cm}{%
      \includegraphics[width=\linewidth]{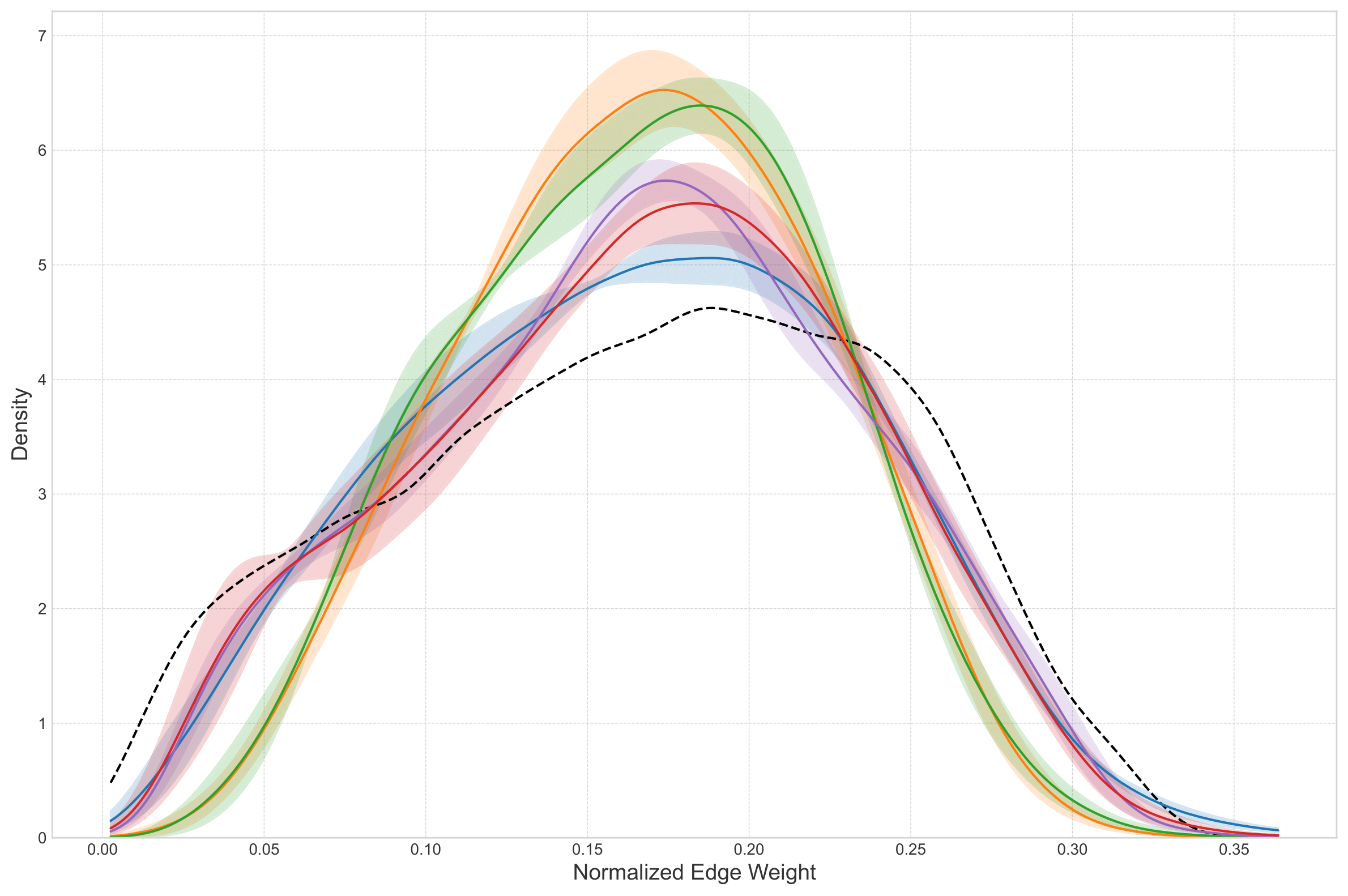}
    }
    \caption*{(d) Edge Weight Distribution}
  \end{minipage}

  \begin{center}
      \vspace{0.3cm}
      \includegraphics[width=0.55\textwidth]{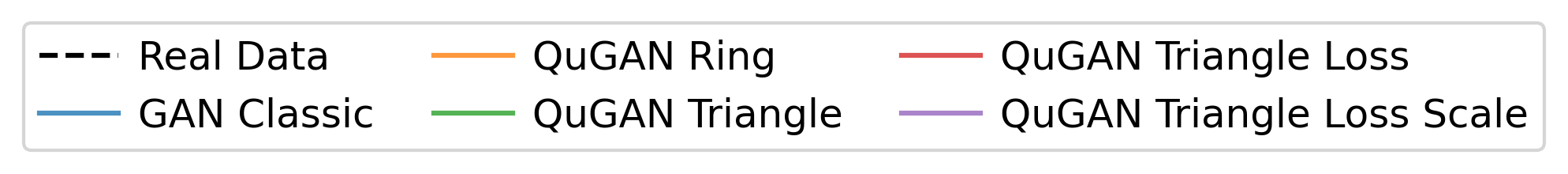}

  \end{center}
  
  \caption{Comparison of model performance on key metrics: (a) edge weight standard deviation, (b) triangle validity , (c) Ptolemaic validity, and (d) edge weight distribution.}
  \label{fig:model_eval_overview}
\end{figure*}

Starting from the \textsc{QuGAN~Triangle} baseline, adding the \emph{variance matching} term (\textsc{QuGAN~Triangle~Loss}) substantially improves statistical fidelity (Wass. $0.009$; JS $0.137$) while leaving global geometric consistency essentially unchanged (4PCM $89.87\%$). Local consistency weakens modestly (TVS $73.42\%$), indicating a measurable but limited trade‑off between the distribution-based (statistics-based) metrics and the validity-based (self-quality-based) metrics. Augmenting the model further with the output‑scaling layer (\textsc{QuGAN~Triangle~Loss+Scale}) does not materially change distributional performance, but improves TVS and 4PCM. Thus scaling acts here as a fine calibration to strengthen validity rather than a major driver of distributional change.

Relative to the \textsc{GAN} baseline, the variance‑regularized Triangle models narrow the statistical gap to within striking distance while retaining competitive geometric validity (and exceeding the classical model in TVS). Qualitatively, Fig.~\ref{fig:model_eval_overview} (d) illustrates, that the variance‑regularized models achieve a closer approximation to the target distribution, highlighting the expressive potential of structured entanglement under targeted regularization.
Jointly examining subplots (a)–(c) of Fig.~\ref{fig:model_eval_overview} reveals a consistent trade-off in the \textsc{GAN}, \textsc{QuGAN Ring}, and \textsc{QuGAN Triangle} models: as geometric validity increases, the standard deviation of the generated distributions decreases. This can be explained by the fact that it becomes more difficult to comply with the constraints when the variance of the edges increases. However, the \emph{variance matching} term can be used to influence the trade-off, while the scaling layer appears to enable a better performance with regard to geometric constraints, two valuable tools for tailoring models to their specific purpose. Our results indicate that the observed under-dispersion is not a insuperable limitation of the circuit itself, but can be compensated through adaption in the training setting.

\subsection{Limitations \& Future Work}
While our results validate topology-guided quantum architectures for structured generative tasks, several limitations remain. The most immediate challenge is scalability. Our Triangle-based circuit requires one qubit per edge and three CNOT gates per triangle. For a complete graph \(K_n\), this implies a quadratic growth in qubit count (\(\mathcal{O}(n^2)\)) and cubic scaling in CNOTs per layer (\(\mathcal{O}(n^3)\)), driven by the \(\binom{n}{3}\) triangles in \(K_n\). 
Though this improves over dense all-to-all schemes (\(\mathcal{O}(n^4)\)), it remains significantly more resource-intensive than hardware-efficient baselines with quadratic growth. As a result, scaling to larger graphs will require more efficient triangle-encoding schemes or approximation strategies, such as sparse or hierarchical entanglement layouts, that preserve geometric priors while mitigating gate overhead.

Moreover, the high entanglement depth and gate count of our best-performing models may be vulnerable to noise on current quantum hardware, which was ignored in our simulations. Future work should assess the robustness of geometry-aware QuGANs under noise, building on prior results that suggest stability~\cite{anand2021noisequgan}.

Finally, our approach relies on handcrafted architectural priors derived from known geometric symmetries. A promising direction is to automate this process, e.g., by learning task-specific entanglement patterns or incorporating differentiable geometric constraints into the training objective. These advances would broaden the applicability of inductive-bias-driven quantum modelling beyond fixed, domain-specific designs.

\section{CONCLUSIONS}
\label{sec:conclusion}
This work demonstrates that quantum generative models can benefit substantially from incorporating domain-specific inductive biases. By aligning the entanglement topology with the intrinsic geometry of the task, we show that performance gains arise not from maximizing entanglement capacity, but from embedding structure-aware priors directly into the quantum circuit.
Further analysis reveals a consistent trade-off: CNOT-based variants enhance geometric consistency, while CRot-based circuits improve statistical fidelity. These effects are robust and highlight the value of architectural choices that respect domain constraints. In addition, a variance matching term and a learnable output-scaling layer boost distributional performance and enable modelling of complex tasks, including high-variance and multimodal distributions. With these refinements, QuGANs closely match classical GANs in fidelity while preserving physically meaningful structures.
Our contributions also extend to a novel application setting: generating weighted networks subject to geometric constraints. This formulation enriches the standard generative modelling task by adding a structural validity criterion, effectively elevating it from learning unconstrained distributions to modelling distributions over implicit geometrically consistent objects. 
Nevertheless, limitations remain. The Triangle ansatz scales cubically in depth with graph size (\(\mathcal{O}(n^3)\)), which challenges near-term applicability, and its behaviour under quantum noise is yet to be explored. More broadly, our findings support a central hypothesis: the true advantage of quantum models lies not in its brute-force capacity, but in their ability to encode domain knowledge natively into quantum structure. This supports a shift from generic design to precision-engineered, problem-aware quantum architectures, using entanglement not as a universal fix, but as a targeted and useful modelling tool.

\section*{ACKNOWLEDGEMENTS}
This paper was partially funded by the German Federal Ministry of Education and Research through the funding program “quantum technologies -- from basic research to market” (contract number: 13N16196).

\bibliographystyle{apalike}
{\small
\bibliography{example}}

@article{preskill2012quantum,
  title={Quantum computing and the entanglement frontier},
  author={Preskill, John},
  journal={arXiv preprint arXiv:1203.5813},
  year={2012}
}

@inproceedings{shor1994algorithms,
  title={Algorithms for quantum computation: discrete logarithms and factoring},
  author={Shor, Peter W},
  booktitle={Proceedings 35th annual symposium on foundations of computer science},
  pages={124--134},
  year={1994}
}

@inproceedings{grover1996fast,
  title={A fast quantum mechanical algorithm for database search},
  author={Grover, Lov K},
  booktitle={Proceedings of the twenty-eighth annual ACM symposium on Theory of computing},
  pages={212--219},
  year={1996}
}

@article{schuld2022quantum,
  title={Is quantum advantage the right goal for quantum machine learning?},
  author={Schuld, Maria and Killoran, Nathan},
  journal={Prx Quantum},
  volume={3},
  number={3},
  pages={030101},
  year={2022},
  publisher={APS}
}

@article{bowles2024better,
  title={Better than classical? the subtle art of benchmarking quantum machine learning models},
  author={Bowles, Joseph and Ahmed, Shahnawaz and Schuld, Maria},
  journal={arXiv preprint arXiv:2403.07059},
  year={2024}
}

@article{rudolph2024trainability,
  title={Trainability barriers and opportunities in quantum generative modeling},
  author={Rudolph, Manuel S and Lerch, Sacha and Thanasilp, Supanut and Kiss, Oriel and Shaya, Oxana and Vallecorsa, Sofia and Grossi, Michele and Holmes, Zo{\"e}},
  journal={npj Quantum Information},
  volume={10},
  number={1},
  pages={116},
  year={2024},
  publisher={Nature Publishing Group UK London}
}

@inproceedings{you2018graphrnn,
  title     = {GraphRNN: Generating Realistic Graphs with Deep Auto-Regressive Models},
  author    = {Jiaxuan You and Rex Ying and Xiang Ren and William L. Hamilton and Jure Leskovec},
  booktitle = {Proceedings of the 35th International Conference on Machine Learning},
  series    = {Proceedings of Machine Learning Research},
  volume    = {80},
  pages     = {5708--5717},
  year      = {2018},
  url       = {https://proceedings.mlr.press/v80/you18a.html}
}

@inproceedings{zhu2022graphsurvey,
  title     = {A Survey on Deep Graph Generation: Methods and Applications},
  author    = {Yujun Zhu and Yihong Du and Yulong Wang and Yanghua Xu and Jing Zhang and Qi Liu and Shu Wu},
  booktitle = {Learning on Graphs Conference},
  publisher = {PMLR},
  year      = {2022},
  pages     = {47--1}
}

@article{dallaire2018quantum,
  title   = {Quantum Generative Adversarial Networks},
  author  = {Pierre-Luc Dallaire-Demers and Nathan Killoran},
  journal = {Physical Review A},
  volume  = {98},
  number  = {1},
  pages   = {012324},
  year    = {2018}
}

@article{lloyd2018quantum,
  title   = {Quantum Generative Adversarial Learning},
  author  = {Seth Lloyd and Christian Weedbrook},
  journal = {Physical Review Letters},
  volume  = {121},
  number  = {4},
  pages   = {040502},
  year    = {2018}
}

@inproceedings{chakrabarti2019qwgan,
  title     = {Quantum Wasserstein Generative Adversarial Networks},
  author    = {Subhankar Chakrabarti and Yiming Huang and Tian Li and Soheil Feizi and Xiaolong Wu},
  booktitle = {Advances in Neural Information Processing Systems},
  volume    = {32},
  year      = {2019}
}

@article{huggins2019tnqgan,
  title   = {Towards Quantum Machine Learning with Tensor Networks},
  author  = {William Huggins and Pranav Patil and Brian Mitchell and K. Birgitta Whaley and Edward M. Stoudenmire},
  journal = {Quantum Science and Technology},
  volume  = {4},
  number  = {2},
  pages   = {024001},
  year    = {2019}
}

@inproceedings{stein2021qugan,
  title     = {QuGAN: A Quantum‐State-Fidelity-Based Generative Adversarial Network},
  author    = {Shijun A. Stein and Bahador Baheri and Dong Chen and Yuhang Mao and Qiang Guan and Ang Li and Bo Fang and Song Fu},
  booktitle = {IEEE International Conference on Quantum Computing and Engineering (QCE)},
  pages     = {71--81},
  year      = {2021}
}

@inproceedings{Shi2020GraphAF,
title={GraphAF: a Flow-based Autoregressive Model for Molecular Graph Generation},
author={Chence Shi and Minkai Xu and Zhaocheng Zhu and Weinan Zhang and Ming Zhang and Jian Tang},
booktitle={International Conference on Learning Representations},
year={2020},
url={https://openreview.net/forum?id=S1esMkHYPr}
}

@inproceedings{zang2020moflow,
  title={Moflow: an invertible flow model for generating molecular graphs},
  author={Zang, Chengxi and Wang, Fei},
  booktitle={Proceedings of the 26th ACM SIGKDD international conference on knowledge discovery \& data mining},
  pages={617--626},
  year={2020}
}

@article{kao2023qganchem,
  title   = {Exploring the Advantages of Quantum Generative Adversarial Networks in Generative Chemistry},
  author  = {Po-Yu Kao and Yi-Chun Yang and Wei-Yuan Chiang and Jung-Yi Hsiao and Yuan Cao and Artem A. Aliper and Fan Ren and Alán Aspuru-Guzik and Alex Zhavoronkov and Min-Hsiu Hsieh},
  journal = {Journal of Chemical Information and Modeling},
  volume  = {63},
  number  = {11},
  pages   = {3307--3318},
  year    = {2023}
}

@article{anand2021noisequgan,
  title   = {Noise Robustness and Experimental Demonstration of a Quantum Generative Adversarial Network for Continuous Distributions},
  author  = {Abhinav Anand and Jonathan Romero and Michiel Degroote and Alán Aspuru-Guzik},
  journal = {Advanced Quantum Technologies},
  volume  = {4},
  number  = {5},
  pages   = {2000069},
  year    = {2021}
}

@inproceedings{rohe2025searoute,
  author    = {Tobias Rohe and Florian Burger and Michael K{\"o}lle and Sebastian W{\"o}lckert and Maximilian Zorn and Claudia Linnhoff-Popien},
  title     = {Investigating Parameter-Efficiency of Hybrid QuGANs Based on Geometric Properties of Generated Sea Route Graphs},
  booktitle = {Proceedings of the 17th International Conference on Agents and Artificial Intelligence (ICAART 2025) - Volume 1},
  pages     = {724--730},
  year      = {2025},
  doi       = {10.5220/0013350200003890},
}

@article{sim2019expressibility,
  title={Expressibility and entangling capability of parameterized quantum circuits for hybrid quantum-classical algorithms},
  author={Sim, Sukin and Johnson, Peter D. and Aspuru-Guzik, Al{\'a}n},
  journal={Advanced Quantum Technologies},
  volume={2},
  number={7-8},
  pages={1900070},
  year={2019},
  publisher={Wiley},
  doi={10.1002/qute.201900070},
  eprint={1905.10876},
  archivePrefix={arXiv},
  primaryClass={quant-ph},
  url={https://doi.org/10.48550/arXiv.1905.10876}
}

@article{kandala2017hardware,
  title   = {Hardware‐Efficient Variational Quantum Eigensolver for Small Molecules and Quantum Magnets},
  author  = {Abhinav Kandala and Antonio Mezzacapo and Kristan Temme and Maika Takita and Markus Brink and Jerry M. Chow and Jay M. Gambetta},
  journal = {Nature},
  volume  = {549},
  pages   = {242--246},
  year    = {2017}
}

@article{meyer2023exploiting,
  title={Exploiting symmetry in variational quantum machine learning},
  author={Meyer, Johannes Jakob and Mularski, Marian and Gil-Fuster, Elies and Mele, Antonio Anna and Arzani, Francesco and Wilms, Alissa and Eisert, Jens},
  journal={PRX quantum},
  volume={4},
  number={1},
  pages={010328},
  year={2023},
  publisher={APS}
}

@article{bako2024problem,
  title={Problem-informed Graphical Quantum Generative Learning},
  author={Bak{\'o}, Bence and Kallus, Zs{\'o}fia and Zimboras, Zoltan},
  journal={Bulletin of the American Physical Society},
  year={2024},
  publisher={APS}
}

@misc{openflights2017,
  author       = {{OpenFlights}},
  title        = {OpenFlights: Airport, airline and route data},
  year         = {2017},
  note         = {Accessed July 7, 2025},
  url          = {https://openflights.org/data.html}
}

@article{sinnott1984haversine,
  author  = {Sinnott, Roger W.},
  title   = {Virtues of the Haversine},
  journal = {Sky \& Telescope},
  year    = {1984},
  volume  = {68},
  number  = {2},
  pages   = {158--159},
  note    = {Bibcode: 1984S\&T....68R.158S}
}

@article{paszke2019pytorch,
  author  = {Adam Paszke and Sam Gross and Francisco Massa and Adam Lerer
             and James Bradbury and Gregory Chanan and Trevor Killeen
             and Zeming Lin and Natalia Gimelshein and Luca Antiga
             and Alban Desmaison and Andreas K{\"o}pf and Edward Z. Yang
             and Zach DeVito and Martin Raison and Alykhan Tejani
             and Sasank Chilamkurthy and Benoit Steiner and Lu Fang
             and Junjie Bai and Soumith Chintala},
  title   = {PyTorch: An Imperative Style, High-Performance Deep Learning Library},
  journal = {CoRR},
  volume  = {abs/1912.01703},
  year    = {2019},
  eprinttype = {arXiv},
  eprint  = {1912.01703},
  url     = {http://arxiv.org/abs/1912.01703}
}

@article{bergholm2018pennylane,
  author  = {Ville Bergholm and Josh A. Izaac and Maria Schuld
             and Christian Gogolin and Nathan Killoran},
  title   = {PennyLane: Automatic Differentiation of Hybrid Quantum-Classical Computations},
  journal = {CoRR},
  volume  = {abs/1811.04968},
  year    = {2018},
  eprinttype = {arXiv},
  eprint  = {1811.04968},
  url     = {http://arxiv.org/abs/1811.04968}
}

@book{villani2009optimal,
  author    = {C{\'e}dric Villani},
  title     = {Optimal Transport: Old and New},
  series    = {Grundlehren der mathematischen Wissenschaften},
  volume    = {338},
  publisher = {Springer},
  address   = {Berlin, Heidelberg},
  year      = {2009},
  doi       = {10.1007/978-3-540-71050-9}
}

@article{lin1991divergence,
  author  = {Jianhua Lin},
  title   = {Divergence Measures Based on the Shannon Entropy},
  journal = {IEEE Transactions on Information Theory},
  volume  = {37},
  number  = {1},
  pages   = {145--151},
  year    = {1991},
  doi     = {10.1109/18.61115}
}

@article{zoufal2019quantum,
  title={Quantum generative adversarial networks for learning and loading random distributions},
  author={Zoufal, Christa and Lucchi, Aur{\'e}lien and Woerner, Stefan},
  journal={npj Quantum Information},
  volume={5},
  number={1},
  pages={103},
  year={2019},
  publisher={Nature Publishing Group UK London}
}

\section*{Technical Appendix}

\begin{figure*}[t]
  \centering
  \begin{minipage}[t]{0.48\textwidth}
    \centering
    \includegraphics[width=\linewidth]{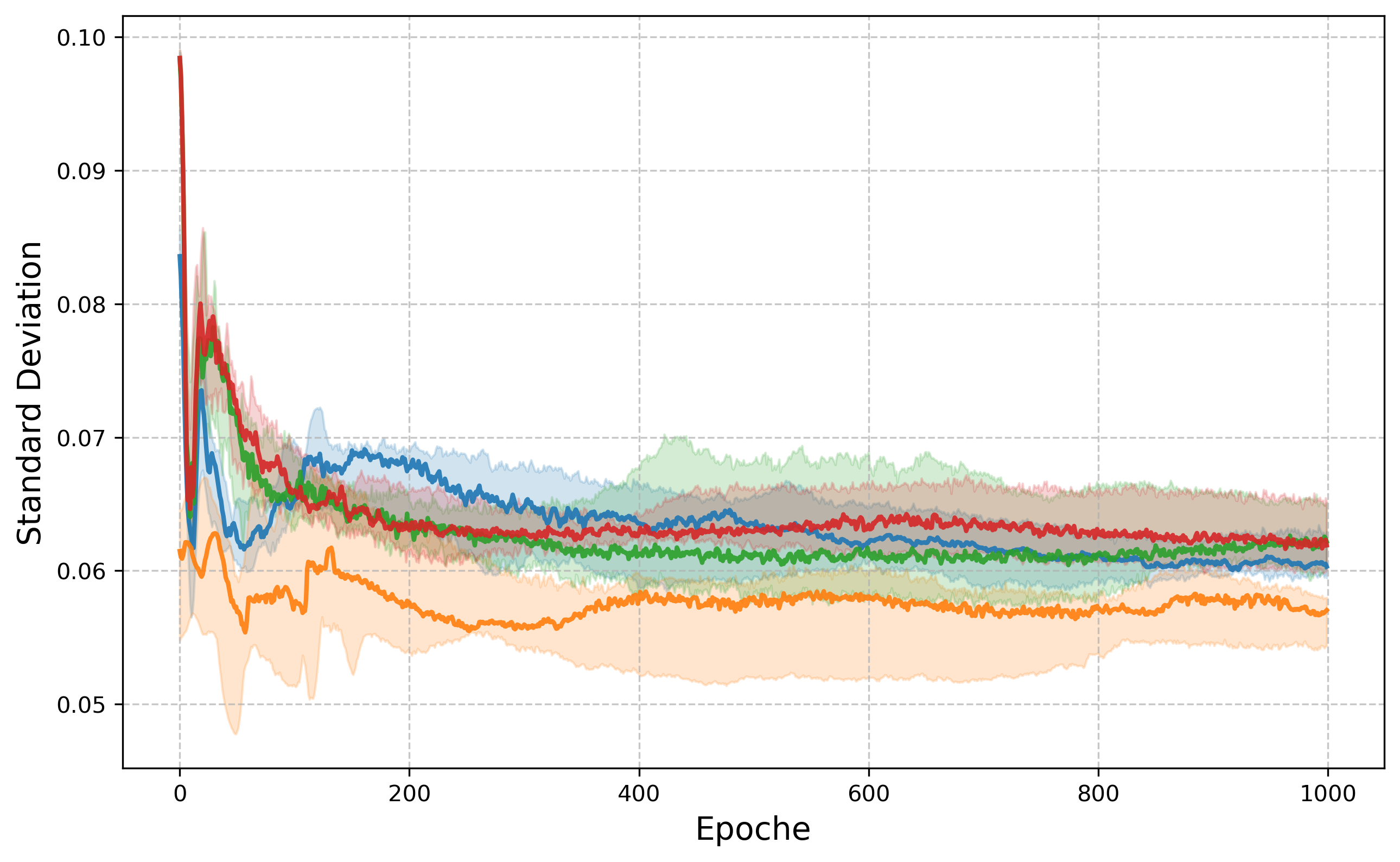}
    \caption*{(a) Standard Deviation over Epochs}
  \end{minipage}
  \hfill
  \begin{minipage}[t]{0.48\textwidth}
    \centering
    \includegraphics[width=\linewidth]{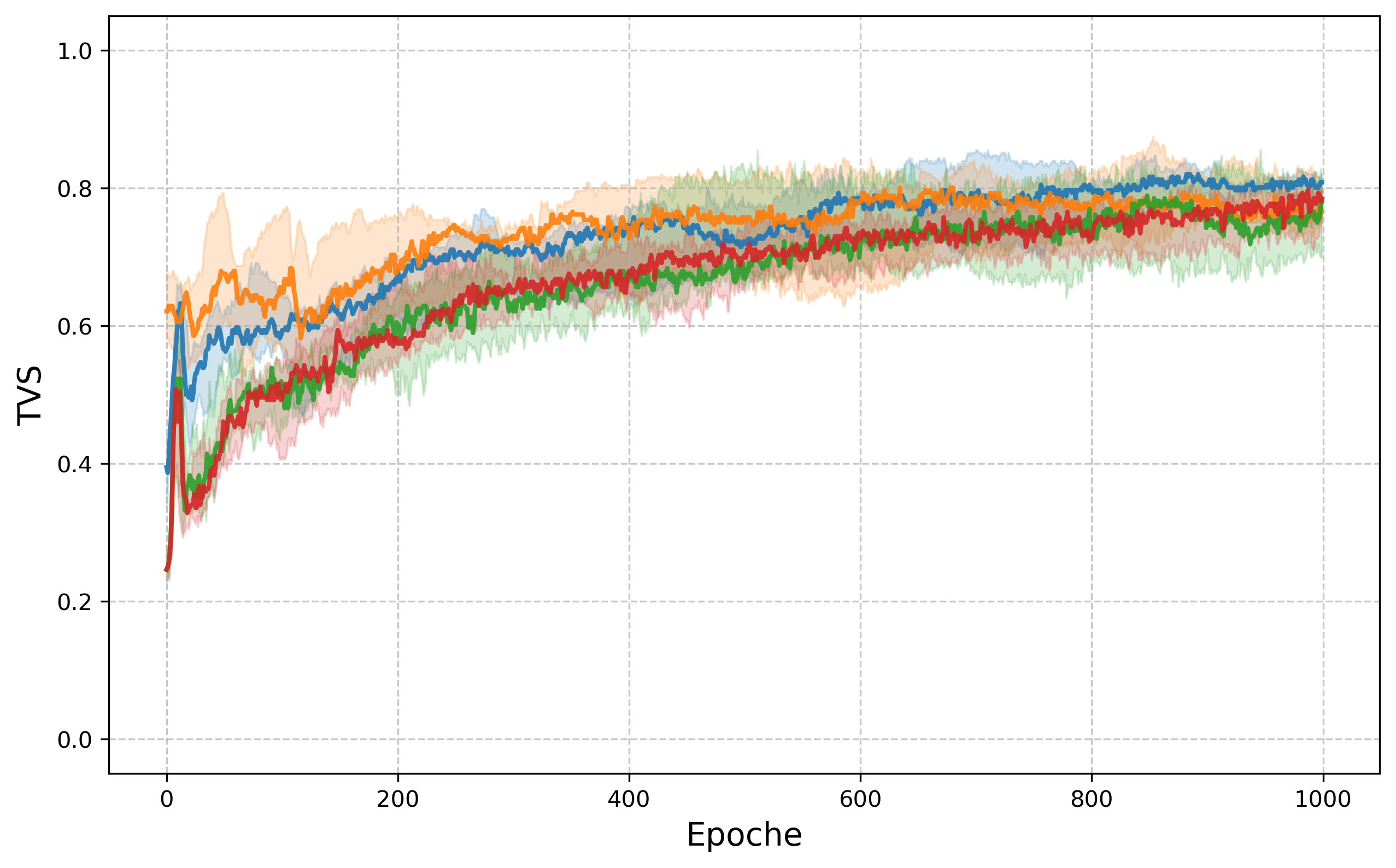}
    \caption*{(b) Triangle Validity over Epochs}
  \end{minipage}
  \vspace{0.5cm}

  \begin{minipage}[t]{0.48\textwidth}
    \centering
    \includegraphics[width=\linewidth]{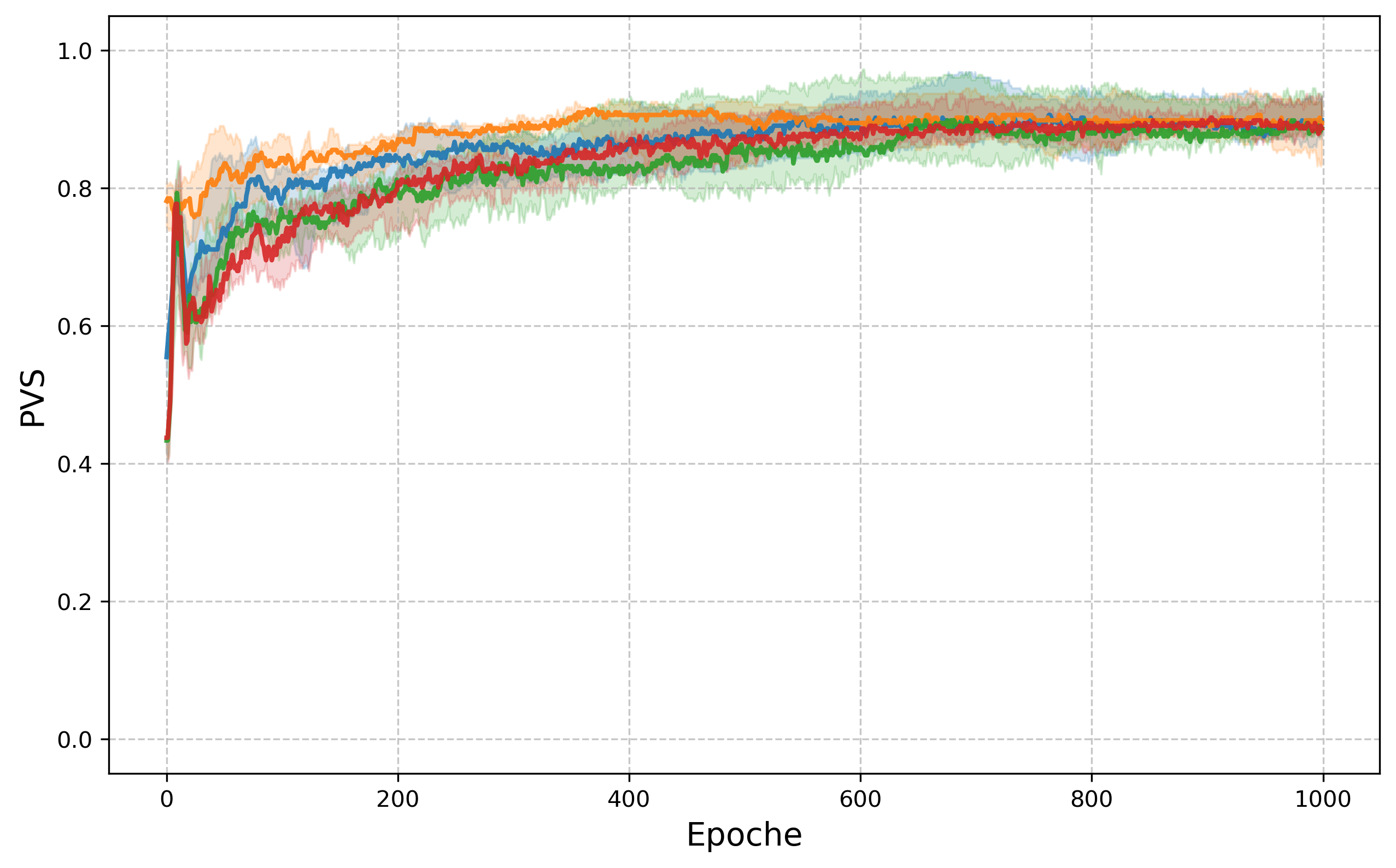}
    \caption*{(c) Ptolemaic Validity over Epochs}
  \end{minipage}
  \hfill
  \begin{minipage}[t]{0.48\textwidth}
    \centering
    \raisebox{0.075cm}{%
      \includegraphics[width=\linewidth]{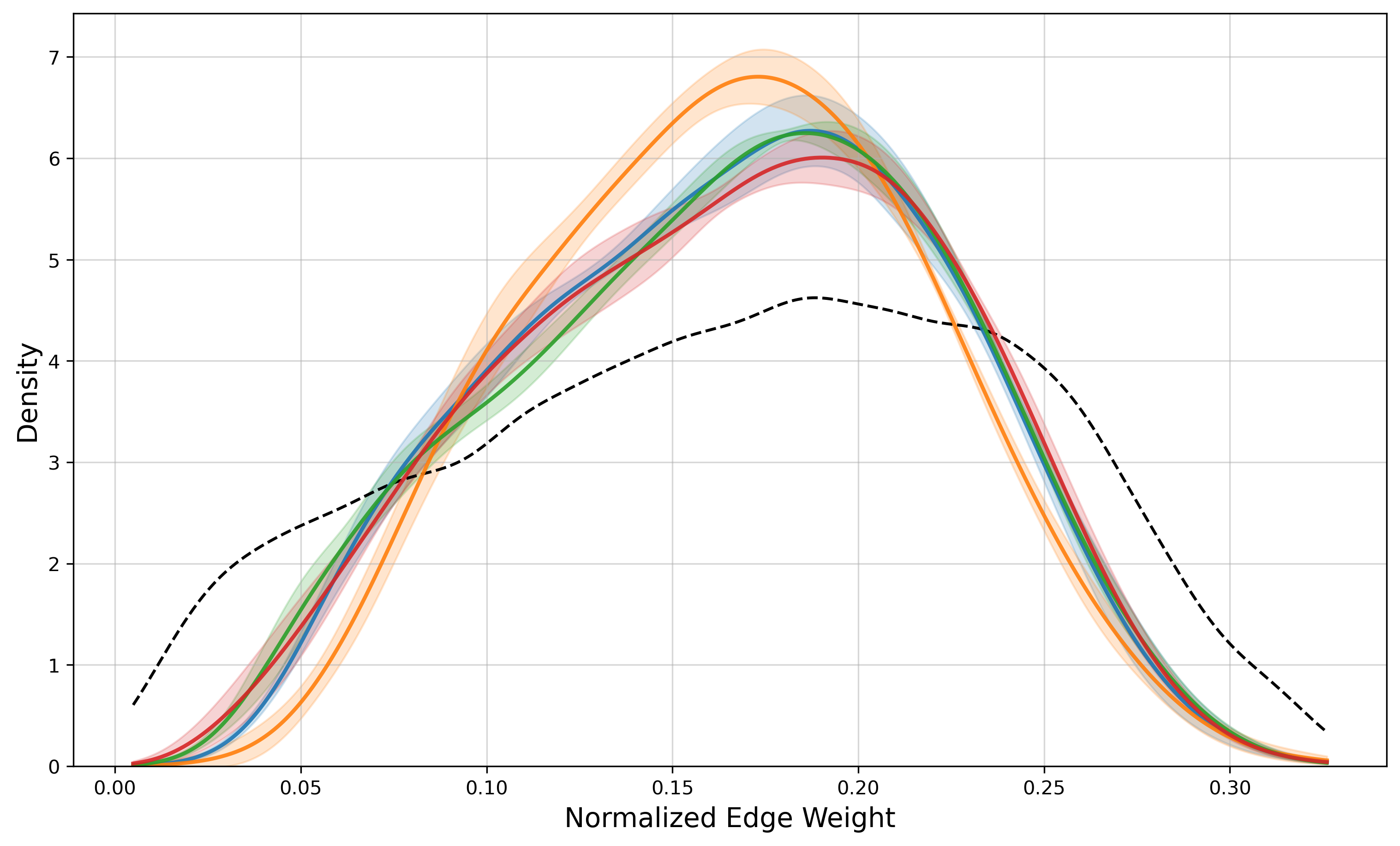}
    }
    \caption*{(d) Edge Weight Distribution}
  \end{minipage}
\begin{center}
    \vspace{0.3cm}
    \includegraphics[width=0.5\textwidth]{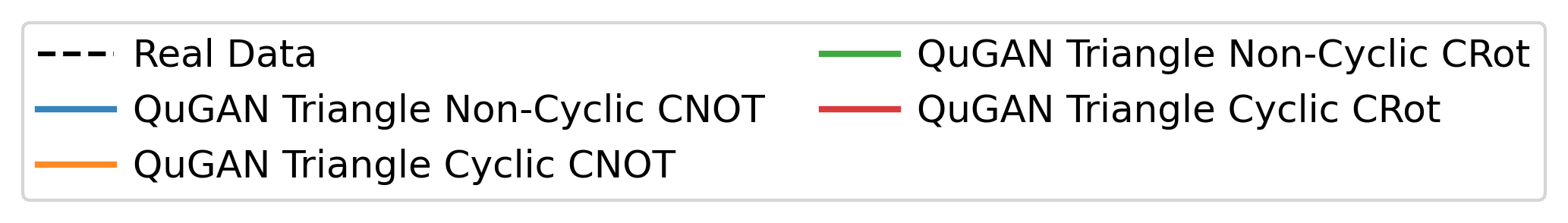}
\end{center}

  \caption{Comparison of model performance on key metrics: (a) edge weight standard deviation, (b) triangle validity , (c) Ptolemaic validity, and (d) edge weight distribution.}
  \label{fig:model_eval_overview_triangle_Variants}
\end{figure*}

\subsection*{A. Analysis Triangle Topology Variants}
Looking at the performance of the Triangle Topology Variants Fig.~\ref{fig:model_eval_overview_triangle_Variants} a. shows that standard deviations are initially high during early training epochs but progressively, slowly decline over training. Concurrently, both the Triangle Validity (Fig.~\ref{fig:model_eval_overview_triangle_Variants} b.) and Ptolemaic Validity (Fig.~\ref{fig:model_eval_overview_triangle_Variants} c.) increase steadily throughout training. The Triangle Validity converges to values between $0.7$ and $0.8$, while Ptolemaic Validity stabilizes slightly below $0.9$. This training dynamic is consistent with expectations: as learning progresses, the QuGAN appears to internalize the geometric constraints inherent in the training data.

Examining the resulting edge-weight distributions, we observe light left-skewness but no evidence of multimodality. As in previous results, the generated distributions still fail to fully replicate the breadth of the real data distribution.

\subsection*{B. The Four‐Point Ptolemaic Consistency Metric (4PCM)}

Generative models for spatial networks must produce graphs that accurately reflect the underlying metric structure of the domain. In the context of flight‐route networks, where vertices are airports and edge weights encode geodesic distances, a single corrupted edge weight can invalidate shortest‐path computations or render network analyses meaningless. While the classical triangle inequality is a widely employed first sanity check, it only certifies local, three‐point consistency and systematically overlooks many higher‐order geometric artifacts. This appendix provides the theoretical foundation, outlines the combinatorial derivation, and presents the formal definition of the 4PCM, which serves as our primary tool for assessing the geometric consistency of generated graphs.

\subsubsection*{Theoretical Foundation: From Unordered Distances to Geometric Constraints}

The core challenge in validating a four‐point graph, such as the complete graph \(K_4\) with vertices \(\{A,B,C,D\}\), is that we are often presented with only an unordered sextuple of its \(\binom{4}{2}=6\) pairwise distances:
\[
\{d_{AB},\,d_{AC},\,d_{AD},\,d_{BC},\,d_{BD},\,d_{CD}\}.
\]
Without knowledge of the graph’s cyclic ordering (e.g., \(A \to B \to C \to D \to A\)), we cannot a priori identify which pairs of edges are adjacent and which are opposite (diagonals). This ambiguity is critical because the primary constraint for four‐point planarity, the Ptolemaic inequality, depends on this distinction. For a quadrilateral \(ABCD\) inscribed in a circle, the Ptolemaic equality states:
\[
d_{AC}\cdot d_{BD} \;=\; d_{AB}\cdot d_{CD} \;+\; d_{AD}\cdot d_{BC},
\]
where \((AC,BD)\) are the diagonals. For general points in a metric space, this becomes an inequality. The 4PCM is designed to resolve the ambiguity of the diagonal pair.

\subsubsection*{Derivation via Perfect Matchings}

To create a test that is independent of any assumed vertex ordering, we consider the combinatorial structure of \(K_4\). A \emph{perfect matching} is a set of edges that covers all four vertices disjointly. In \(K_4\), there are exactly three such matchings, which correspond to the three ways to partition the six edges into pairs of opposite edges:
\[
\{(AC,BD)\},\quad \{(AB,CD)\},\quad \{(AD,BC)\}.
\]
Each of these three matchings defines a potential set of diagonals and yields a corresponding Ptolemaic inequality. The 4PCM is therefore defined by the simultaneous satisfaction of all three possible inequalities:
\begin{align}
\text{(i)}\quad & d_{AC}\,d_{BD} \;\le\; d_{AB}\,d_{CD} \;+\; d_{AD}\,d_{BC},\\
\text{(ii)}\quad & d_{AB}\,d_{CD} \;\le\; d_{AC}\,d_{BD} \;+\; d_{AD}\,d_{BC},\\
\text{(iii)}\quad & d_{AD}\,d_{BC} \;\le\; d_{AC}\,d_{BD} \;+\; d_{AB}\,d_{CD}.
\end{align}
Testing only one of these implicitly assumes a specific matching. For instance, a random sextuple of distances can satisfy inequality (i) by chance, even though no valid cyclic ordering exists. Only when all three inequalities hold simultaneously can we be assured that the six given distances can be realized by four points in a Euclidean plane. This set of conditions is therefore necessary (and, for Euclidean space, also sufficient) for the geometric realizability of the quadruple. While the classical Ptolemaic equality relates to points on a circle, these three inequalities provide a more general test for planarity using only the edge lengths.

\subsubsection*{Practical Application and Empirical Validation}
The 4PCM is designed for operational scenarios in which edge weights are subject to rounding, normalization, and spherical-to-Euclidean projection noise. Under these conditions, it acts as a robust filter for inconsistent samples.

To establish a baseline and select an appropriate tolerance, we applied the 4PCM to our benchmark dataset of 10{,}000 randomly sampled airport quadruples. The results were as follows:

\begin{itemize}
  \item Triangle Inequality: 10{,}000 out of 10{,}000 samples satisfied the constraint (100.00\%).
  \item 4PCM: 9{,}804 out of 10{,}000 samples satisfied the constraint (98.04\%).
\end{itemize}
Based on this analysis, we selected a uniform tolerance of $\varepsilon = 10^{-2}$ for all geometric validity evaluations (both TVS and 4PCM). This threshold reflects a pragmatic balance between numerical precision and robustness to real-world distortions. Specifically, it accounts for (i) rounding errors due to floating-point arithmetic, (ii) the effects of normalization procedures that scale edge weights to sum to one, and (iii) the geometric discrepancy introduced when projecting spherical distances onto a Euclidean space. The fact that 98.04\% of real samples satisfy the 4PCM constraint under this setting suggests that the test is both discriminative and tolerant. It effectively distinguishes geometrically plausible configurations while remaining resilient to measurement artifacts, thereby validating the 4PCM as a practical tool for structure-aware filtering in noisy real-world graph data.

\vspace{0.5em} 
\subsubsection*{Computational Efficiency}
Evaluating the three inequalities for a given quadruple requires only accessing the six pairwise distances and performing a small, constant number of arithmetic operations. This results in an O(1) complexity per quadruple, comparable to that of the simpler triangle inequality test. The significance lies in achieving a more comprehensive four-point geometric consistency check without a substantial increase in computational overhead. This efficiency makes the 4PCM highly suitable for large-scale validation as a post-processing filter, effectively complementing the triangle inequality by scrutinizing higher-order geometric structures.

\end{document}